\documentclass[11pt,twoside]{article}
\usepackage{graphicx}

\usepackage{asp2004}

\begin{document}
\title{HST Stellar Standards with 1\% Accuracy in Absolute Flux}  
\author{Ralph C.\ Bohlin} 
\affil{Space Telescope Science Institute, 3700 San Martin Drive, Baltimore, MD 21218}
\begin{verbatim}
      email: bohlin@stsci.edu
 \end{verbatim}

\begin{abstract}
Free of any atmospheric contamination, the \textit{Hubble Space Telescope} provides the
best available spectrophotometry from the far-UV to the near-IR for stars as
faint as $V\sim16$. The \textit{HST} CALSPEC standard star network is based on
three standard candles: the hot, pure hydrogen white dwarf (WD) stars G191B2B,
GD153, and GD71, which have Hubeny NLTE model flux calculations that require the
atomic physics for only one atom. These model flux distributions are normalized
to the absolute flux for Vega of $3.46\times10^{-9}$~erg cm$^{-2}$ s$^{-1}$~\AA$^{-1}$ 
at 5556~\AA\ using precise Landolt $V$~band photometry and the $V$~bandpass 
function corrected for atmospheric transmission by M.~Cohen. The three
primary WD standards provide absolute flux calibrations for FOS, STIS, and
NICMOS spectrophotometry from these instruments on the \textit{HST}. About
32~stellar spectral energy distributions (SEDs) have been constructed with a
primary pedigree from the STIS data, which extends from 1150~\AA\ for the hot
stars to a long wavelength limit of 1~$\mu$m. NICMOS grism spectrophotometry
provides an extension to 1.9~$\mu$m in the IR for 17 of the \textit{HST} standards and
longward to 2.5~$\mu$m for a few of the brighter stars. Included among these \textit{HST}
standards are Vega, the Sloan standard BD+17$^{\circ}$4708, three bright solar
analog candidates, three cool stars of type M or later, and five hot WDs. In
addition, four K~giants and four main sequence A-stars have NICMOS
spectrophotometry from 0.8--2.5~$\mu$m. The WD fluxes are compared to their
modeled SEDs and demonstrate an internal precision of 1--2\%, while the A-stars
agree with the Cohen IR fluxes to $\sim$2\%. Three solar analog candidate stars
differ from the solar spectrum by up to 10\% in the region of heavy line
blanketing from 3000--4000~\AA\ and show differences in shape of $\sim$5\% in
the IR around 1.8~$\mu$m.
\end{abstract}

\section{Introduction}

To establish a set of stellar flux standards, one or more stars must be
established with known absolute spectral energy distributions (SEDs). One method
for establishing these so-called ``standard candles'' in the sky is to determine
the composition, effective temperature, and surface gravity and then compute a
model atmosphere SED. The simplest case with the best known atomic physics is
for a hot pure hydrogen white dwarf (WD) star, where the temperature and gravity
are determined from fits of model line profiles to the observed Balmer lines
(e.g., Finley, Koester, \& Basri 1997, Barstow et~al.\ 2003). These relative SEDs
become absolute when normalized to the measured flux for Vega using precise 
$V$~band photometry (e.g., Landolt 1992). Bohlin (2003) switched from LTE to NLTE for
calculating the model flux distributions with the Hubeny TLUSTY code (Hubeny \&
Lanz 1995).

\begin{figure}
\centering
\includegraphics[width=\textwidth]{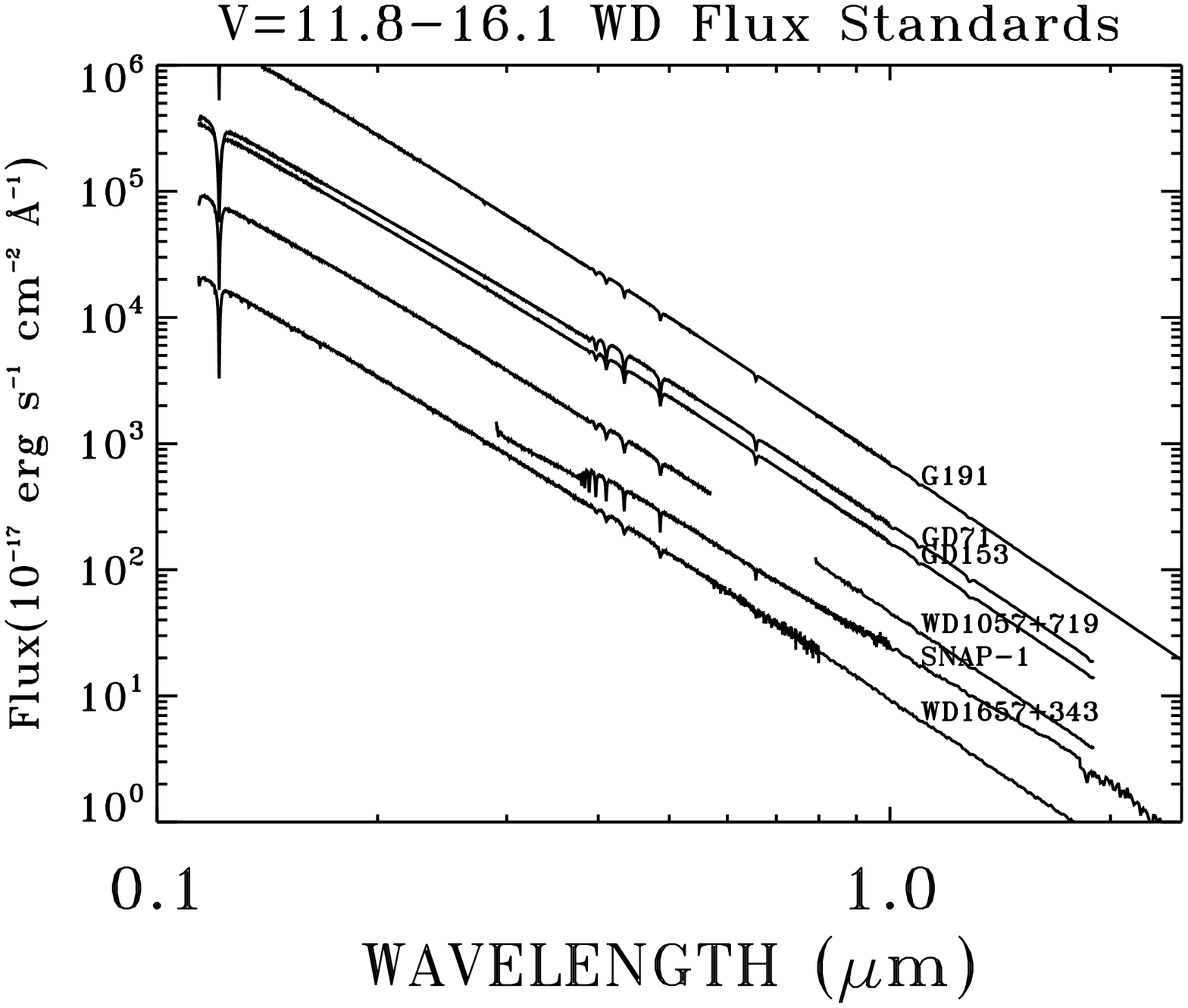}
\caption{STIS or FOS fluxes below 1~$\mu$m and NICMOS above 0.8~$\mu$m. The
noise level is evident in the fainter stars near the STIS long wavelength limit
of 1~$\mu$m and in the NICMOS spectrum of the faint SNAP-1 beyond 1.9~$\mu$m.}
\end{figure}

Following the establishment of the absolute SEDs for the primary stars, a flux
distribution for another star is established by the response relative to the
primaries when observed by a linear spectrometer of constant sensitivity. In
practice, no such instrumentation exists, especially for ground-based
observations through an atmosphere with a transparency that can vary on short
time scales. Even in space, focus variations, sensitivity degradation with time,
and non-linear detectors limit the precision of the transfer of the
calibration from the primary standards to bright secondary standard stars. For
fainter stars photon, statistics also contributes to the uncertainties.

From its installation in 1997 until its death in 2004, the low dispersion
($R\sim1000$) modes of STIS were the premier instrumentation for establishing
flux standards from 1150--10200~\AA\  (Bohlin 2000; Bohlin, Dickinson, \&
Calzetti 2001). Considerable effort has been expended in tracking the STIS
changes in sensitivity with time (Stys, Bohlin, \& Goudfrooij 2004 and
unpublished updates) and in correcting the CCD observations for loss of charge
transfer efficiency (CTE; Goudfooij \& Bohlin 2006). In the 0.8--2.5~$\mu$m
range, the NICMOS $R\sim200$ grism spectrophotometry has been calibrated and the
count rate dependent non-linearity has been characterized by Bohlin, Lindler, \&
Riess (2005) and Bohlin, Riess, \& de~Jong (2006).

\begin{figure}
\centering
\includegraphics[angle=90,width=.89\textwidth]{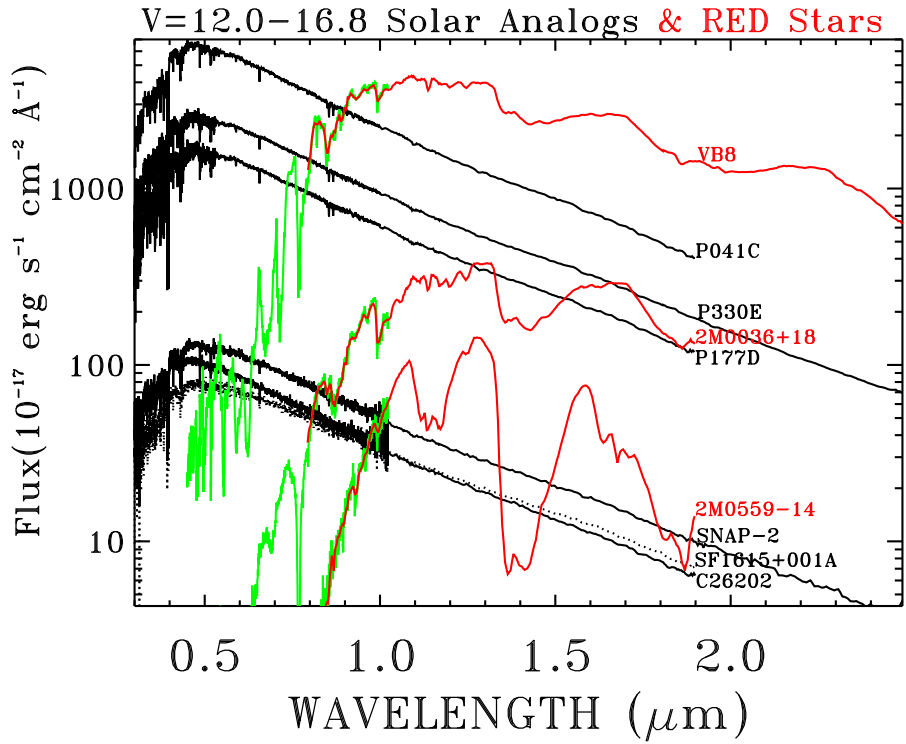}
\caption{CALSPEC stellar flux distributions as in Figure~1. The lower resolution
of the NICMOS observations (red) is evident where the cool star spectra overlap
with STIS (green). The solar analog SF1615+001A is dotted for clarity.}
\end{figure}

Figures 1--2 include most of the SEDs of the stars with NICMOS observations.
Figure~1 shows the hotter WD stars, while Figure~2 illustrates the solar analog
and cooler stars. The wavelength coverage, the spectral types, and the dynamic
range are typical of the standard observational capabilities of the \textit{HST}
spectrographs. Because of the highly over subscribed demand for \textit{HST} time, sky
coverage of the \textit{HST} CALSPEC standard star network is minimal.\footnote{The absolute
spectral energy distributions discussed in this paper are available in digital
form at http://www.stsci.edu/hst/observatory/cdbs/calspec.html.}

\section{White Dwarf Stars}
\begin{figure}
\centering
\includegraphics[width=.85\textwidth]{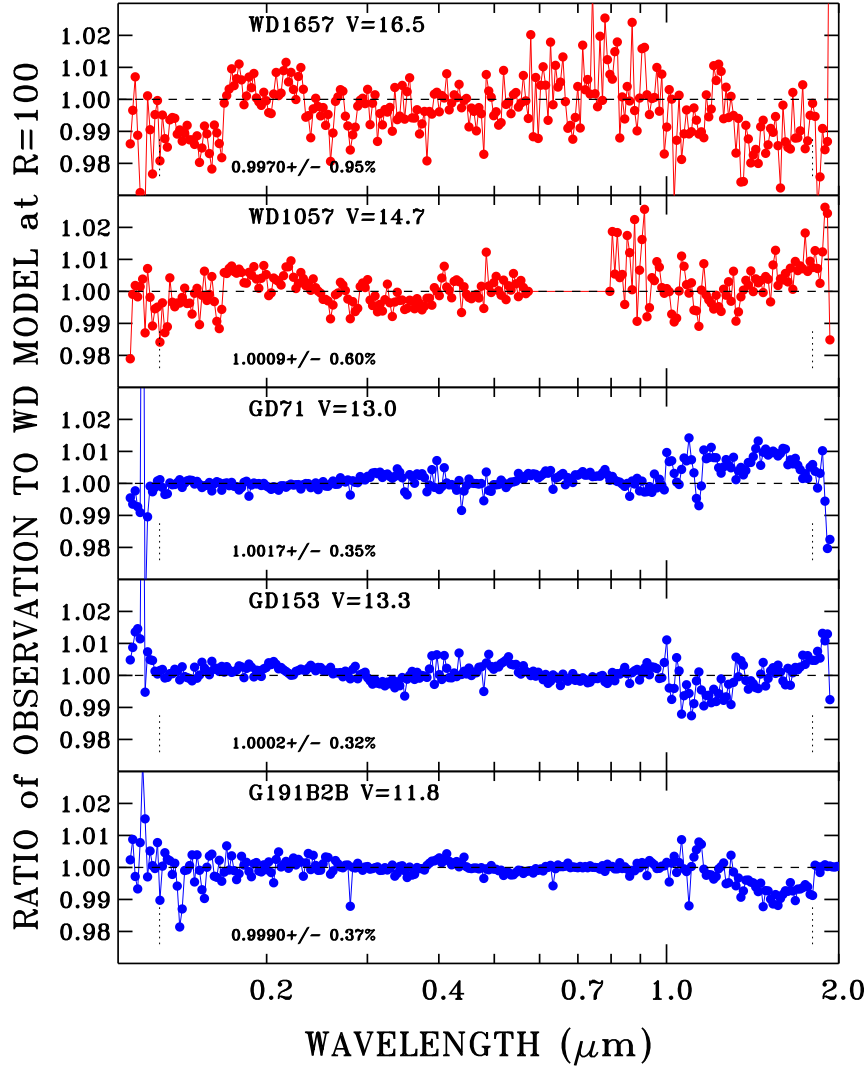}
\caption{Ratio of the measured to predicted flux at a resolution of $R=100$ for
the five stars with pure hydrogen NLTE models. STIS observations are used below
1~$\mu$m and NICMOS spectra are used above 1~$\mu$m, except for the bright
G191B2B where FOS spectrophotometry is used below 0.385~$\mu$m and except for
the two faint WD stars, where the NICMOS data are used down to 0.8~$\mu$m. No
observations exist for WD1057+719 from 0.57 to 0.8~$\mu$m. Faint vertical dotted
lines at 0.13 and 1.8~$\mu$m delineate the range for the average and \% rms of
the ratio that is written in each panel. The brighter primary standards (blue)
in the bottom three panels define the absolute sensitivities; and the average of
these three sets of STIS or NICMOS ratios should be near unity at any
wavelength. Except at the core of the strong Ly$\alpha$ line, the internal
consistency of the three primary standards is usually better than 0.5\% for the
STIS data and $<$1\% for NICMOS. The fainter and less well observed secondary
standards (red) in the top two panels have more points with $>$1--2\% errors; but
only the faintest star WD1657+343 has any error greater than 3\%.}
\end{figure}

The $T_\mathrm{eff}$ and $\log g$ for the five WD stars are from Finley,
Koester, \& Basri (1997), who used LTE models to fit the Balmer lines. The switch
to NLTE models with the same $T_\mathrm{eff}$ and $\log g$ values could cause a
systematic error in the \textit{HST} flux scale by as much as 2\% in the IR and up to 4\%
in the far-UV. However, Figure~3 demonstrates internal consistency among the
three primary standards to $<$0.5\% for STIS and to $\sim$1\% for the NICMOS
data, so that any changes to the models must be nearly the same function of
wavelength for all three stars. An effort to determine the proper NLTE
$T_\mathrm{eff}$ and $\log g$ for the primary standards is underway with D.~Finley 
and I.~Hubeny and will be reported separately. This paper concentrates on
a discussion of the transfer of the primary fluxes to the secondary stars. The
fidelity of this instrumental transfer process is independent of any systematic
adjustment of the fluxes that may arise from a revised set of NLTE models for
the three primary standards.

In Figure~3 longward of Ly$\alpha$ from 0.125 to 1~$\mu$m, virtually all of the
continuum points for the STIS primary standards (blue points) are within $\pm$
0.3\% of unity. The points that do exceed 0.3\% are at the hydrogen lines or are
at the positions of the interstellar/circumstellar features, such as C~\textsc{ii} 1334~\AA\ 
or Mg~\textsc{ii} 2800~\AA. G191B2B is too bright to observe with the STIS MAMA
detectors; and the older FOS fluxes are used below 3850~\AA, where a larger
scatter is seen. There are 6 to 23 separate STIS spectra that are averaged at
each wavelength for each of the three primary standards.

Longward of 1~$\mu$m, only two observations of each star have been obtained with
the NICMOS grisms, so that the typical NICMOS repeatability of 1\% dominates the
scatter. The NICMOS observation are dithered over a Y-range of 16 pixels in a
relatively blemish free region in the lower left quadrant of NIC3 (Bohlin,
Lindler, \& Riess 2005). Due to imperfect flat fields, a larger scatter of $\sim$3\%
might be expected for spectra at arbitrary locations on the detector.

\begin{figure}    
\centering
\includegraphics[width=\textwidth]{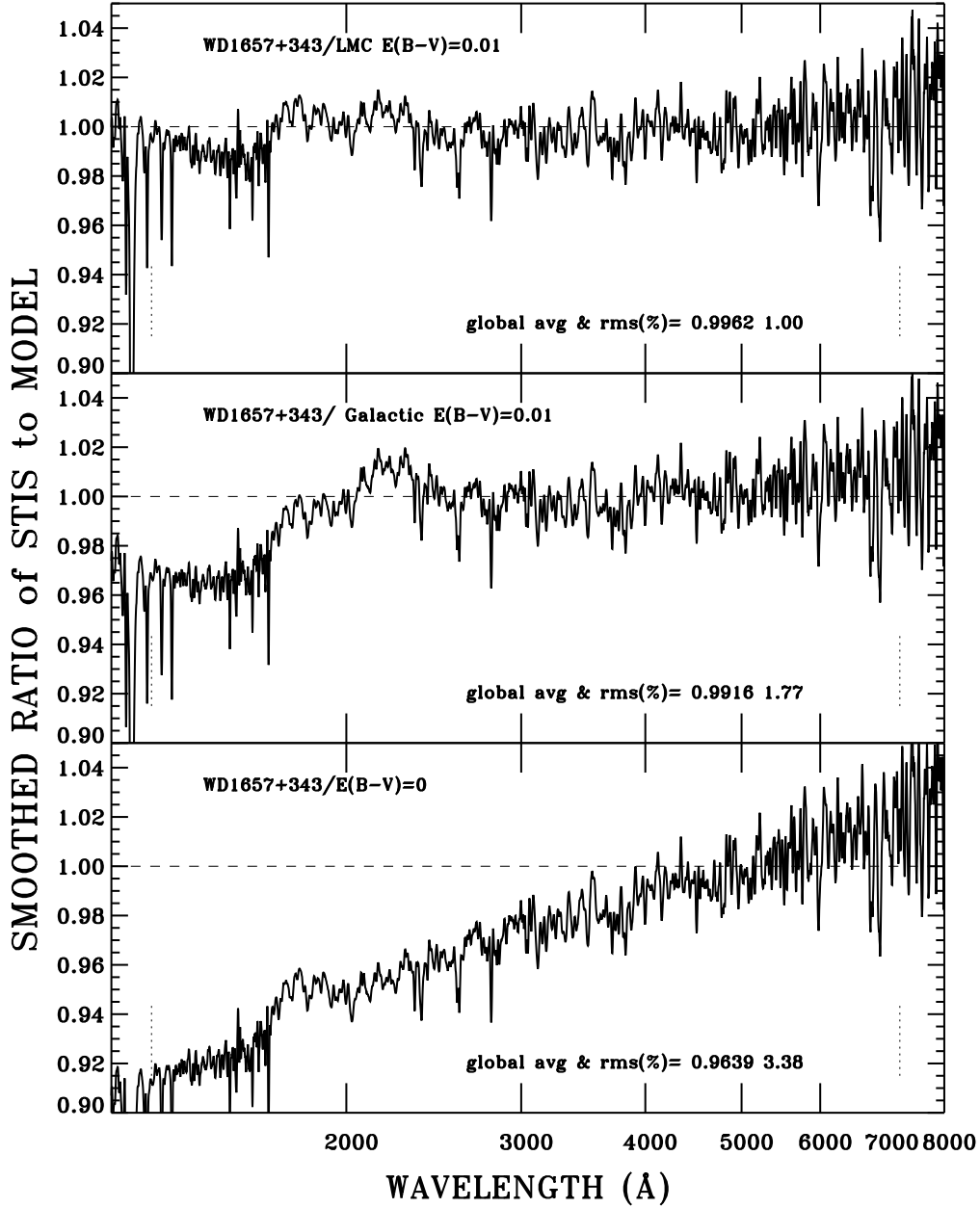}
\caption{Ratio of the measured continuum flux for WD1657+343 to the model
predicted from the Balmer line analysis (lower panel). Reddening the model with
$\mathrm{E(B-V)=0.01}$ improves the fit for the average Milky Way extinction
(middle panel), while reddening with the LMC extinction curve leaves residuals
of only $\sim$1\% (upper panel). There are several narrow interstellar lines
between Ly$\alpha$ at 1216~\AA\ and Mg~\textsc{ii} at 2800~\AA.}
\end{figure}

While the interstellar reddening is negligible for the primary standards, the
fainter secondary standards suffer from a small amount of reddening, as
illustrated in Figure~4 for WD1657+343. The substantially better fit in the
far-UV with the LMC reddening curve than with an average Galactic reddening is a
bit surprising; but such a tiny local amount of dust could be processed
differently than on average in the Galaxy. Similarly for WD1057+719, the model
fits the observed continuum best for an LMC reddening with
$\mathrm{E(B-V)=0.003}$. The ratio of measured fluxes to their reddened models
for these two fainter WD star are shown in Figure~3, as the red points. Because
the residuals for all five WD stars are within measurement uncertainties, the
noise-free, reddened models define the best estimates of the standard star flux
distributions at all wavelengths, except where there are interstellar lines. The
Ly$\alpha$ and Balmer lines of the models cannot be compared precisely to the
observation, because the STIS line-spread function (LSF) has wide wings and is
not well determined.

\section{Vega}

Bohlin \& Gilliland (2004, BG) measured the flux for Vega using STIS spectra
that are heavily saturated, but remarkably linear. Multiple observations repeat
to better than 0.5\%. However, small adjustments to the STIS calibration have
changed the derived absolute spectrophotometry by up to 2\% at the longest
wavelengths, where the improved CTE correction of Goudfooij \& Bohlin (2006) is
the dominate reason for revising the BG results. Below 7000~\AA, flux changes
are $<$0.5\%, eg. the revised V=0.023 $\pm$ 0.008 is 0.003mag brighter in the
Johnson  $V$~band. The new CTE correction changes the slope in the Paschen
continuum and brings the measured slope into agreement with a 9400~K model
($\log g=3.90$, [M/H]~$=-0.5$, and zero microturbulent velocity) from Kurucz
(2005a), as shown in Figure~5. In general, the model fluxes agree with the
observed flux to $\sim$1\% longward of 3200~\AA, except in the Balmer confluence
from $\sim$3700--3900~\AA\ where the atomic physics is most difficult. In one
region from 4200--4700~\AA, the 9550~K Kurucz model adopted by BG fits better by
$\sim$1\%; but on the whole, the 9400~K model is now the best fit, especially
below the temperature sensitive Balmer Jump and at the long wavelengths where
continuity with the model is required to extend the predicted flux beyond the
1~$\mu$m STIS limit. Because of the excellent agreement of the model with the
observations to within the uncertainties, the model is used to define the Vega
standard star flux longward of 5300~\AA.\footnote{The composite Vega spectrum is
a binary fits table named $alpha\_lyr\_stis\_003.fits$ and is available at 
http://www.stsci.edu/hst/observatory/cdbs/calspec.html.}

\begin{figure}
\centering
\includegraphics*[angle=90,width=.98\textwidth]{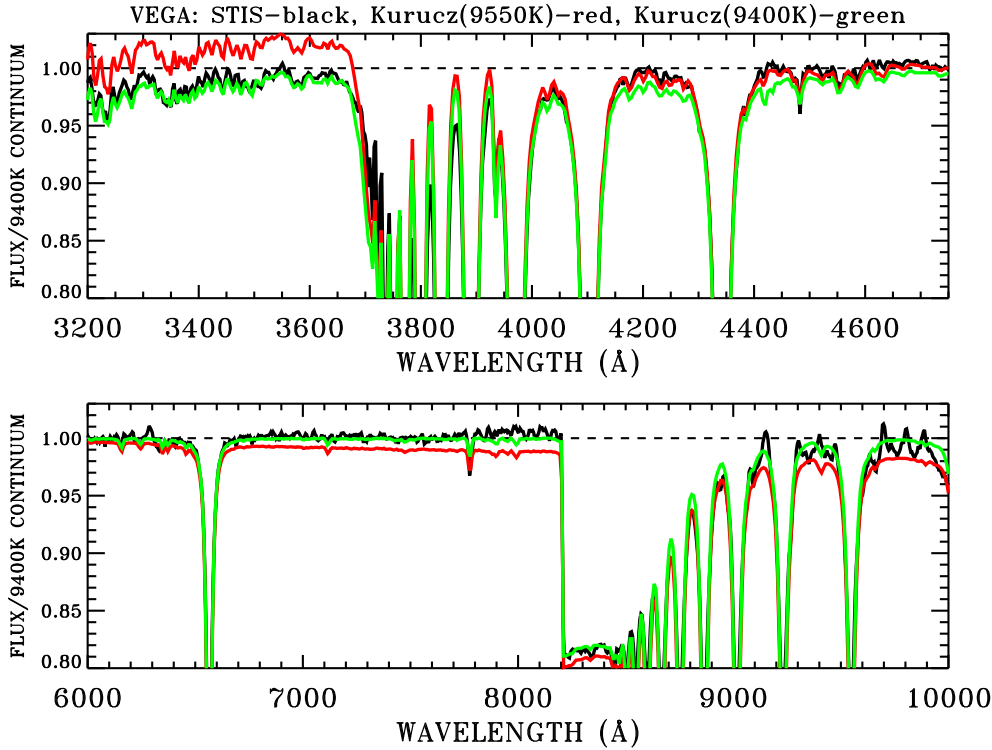}
\caption{Ratio to the 9400~K continuum level for the STIS measured fluxes
(black), the Kurucz 9400~K model (green), and the Kurucz 9550~K model (red). The
models and their continuum levels are all normalized to $3.46\times10^{-9}$~erg
cm$^{-2}$ s$^{-1}$~\AA$^{-1}$ at 5556~\AA\ (Megessier 1995). The observations
agree much better with the cooler model at the shortest and longest wavelengths,
where the effects of the temperature differences are most significant. The
hotter model agrees better with the measured flux by $\sim$1\% at 4200--4700~\AA.}
\end{figure}

The new value of $T_\mathrm{eff}=9400$~K brings the STIS results into agreement
with the previously accepted Vega model that has been the IR standard per the
series of papers by Cohen and co-authors beginning with Cohen et~al.\ (1992). A
new determination of the angular diameter of 3.44~mas for a rapidly rotating,
pole-on model of Vega (Peterson et~al.\ 2006) suggests that the old value of
$3.24\pm0.07$~mas (Hanbury Brown et~al.\ 1974) is wrong. The single temperature
$T_\mathrm{eff}=9400$~K model implies an angular diameter of 3.335~mas, while
9550~K corresponds to 3.273~mas. Aufdenberg, et~al.\ (2006) determine an angular
diameter of 3.33~mas from their rapidly rotating, pole-on model of their
interferometric observations of Vega. The single temperature
$T_\mathrm{eff}=9400$~K Kurucz model fits the STIS SED to 1--2\% from
3200--10000~\AA, even though the rapid rotation produces a temperature range
over the observable surface of 7900--10150~K. The continuum level of the
Aufdenberg multi-temperature model agrees with the Kurucz 9400~K model from
3200--10000~\AA. At longer and shorter wavelengths, a multi-temperature model
may be required to achieve a $\sim$1\% precision.

\section{Solar Analog Stars}
The use of a star of solar spectral type (G2V) as a standard flux candle relies
on the assumption that the solar analog star has the same intrinsic SED as the
sun. However, fundamental limitations to this technique include the facts that
no star is an exact solar analog, the solar SED itself still has uncertainties
of a few percent especially in the IR, and solar variations in the continuum are
greater than 1\% below 2500~\AA\ (e.g., Rottman et~al.\ 2005). Despite the drawbacks
inherent in solar analogs, comparison of the sun to the \textit{HST} spectra of three
candidates stars helps to quantify the limitations of the technique. 

\begin{figure}   
\centering
\includegraphics[angle=90,width=.82\textwidth]{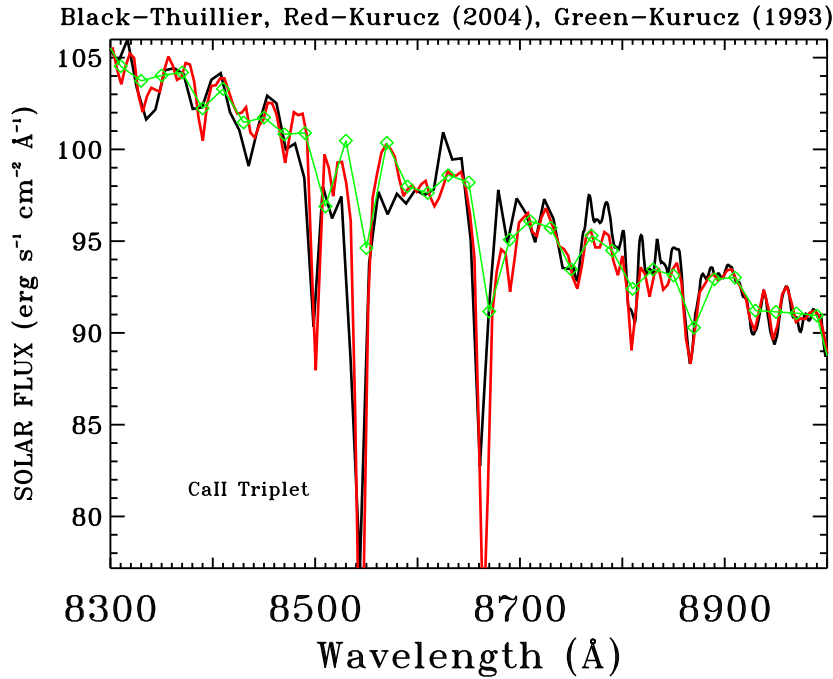}
\caption{A small region of the solar spectrum that includes the strong CaII
triplet. The data of Th03 (black) is compared to a modern 2004 Kurucz
model (red) and to a 1993 vintage Kurucz model (green line and diamonds).}
\end{figure}

Figure~6 compares a small segment (8300--9000~\AA) of the measured solar
spectrum from Thuillier et~al.\ (2003, Th03) with two versions of Kurucz models.
The hi-fidelity $T_\mathrm{eff}=5777$~K model (Kurucz 2004) is a high resolution
computation smoothed to a resolution of $R=1000$, while all of the 1993 vintage
Kurucz models are on a coarser wavelength grid. The observed spectral features
correspond remarkably well to those in the 2004 model, although not always with
exactly the predicted strengths. The green 1993 model tracks the red 2004 model
well in the continuum; but the line strengths are poorly reproduced,
especially for the three strongest lines in Figure~6.

The stars P041C, P177D, and P330E have spectrophotometry from STIS and NICMOS in
the range 0.3--2~$\mu$m, where solar variability is $<$1\%. The compilation of
the solar spectrum from Th03 covers the 0.2--2.4~$\mu$m range at a resolution of
10~\AA. Figure~7 shows a comparison of the measured solar analogs to Th03 in
50~\AA\ bins, where the model is normalized to the measured fluxes at
7000--8000~\AA. Below 4500~\AA, there is considerable structure in the ratio,
while differences are more continuous at the longer wavelengths. These
continuous deviations can be significantly reduced, if small corrections for
interstellar reddening and small deviations from the solar temperature are made.
Bohlin, Dickinson, \& Calzetti (2001) suggested that P177D and P330E are
reddened by E(B-V)~= 0.04 and 0.03, respectively. With a STIS calibration updated
in 2006 and with the NICMOS extension, reddenings of only 0.02 and 0.01 are
required to bring the ratios closer to unity, as shown in Figure~8. The Galactic
average extinction curve of Savage \& Mathis (1979) is used to redden the solar
spectrum for the quotients shown for P177D and P330E in Figure~8.
\begin{figure} 
\centering
\includegraphics[width=\textwidth]{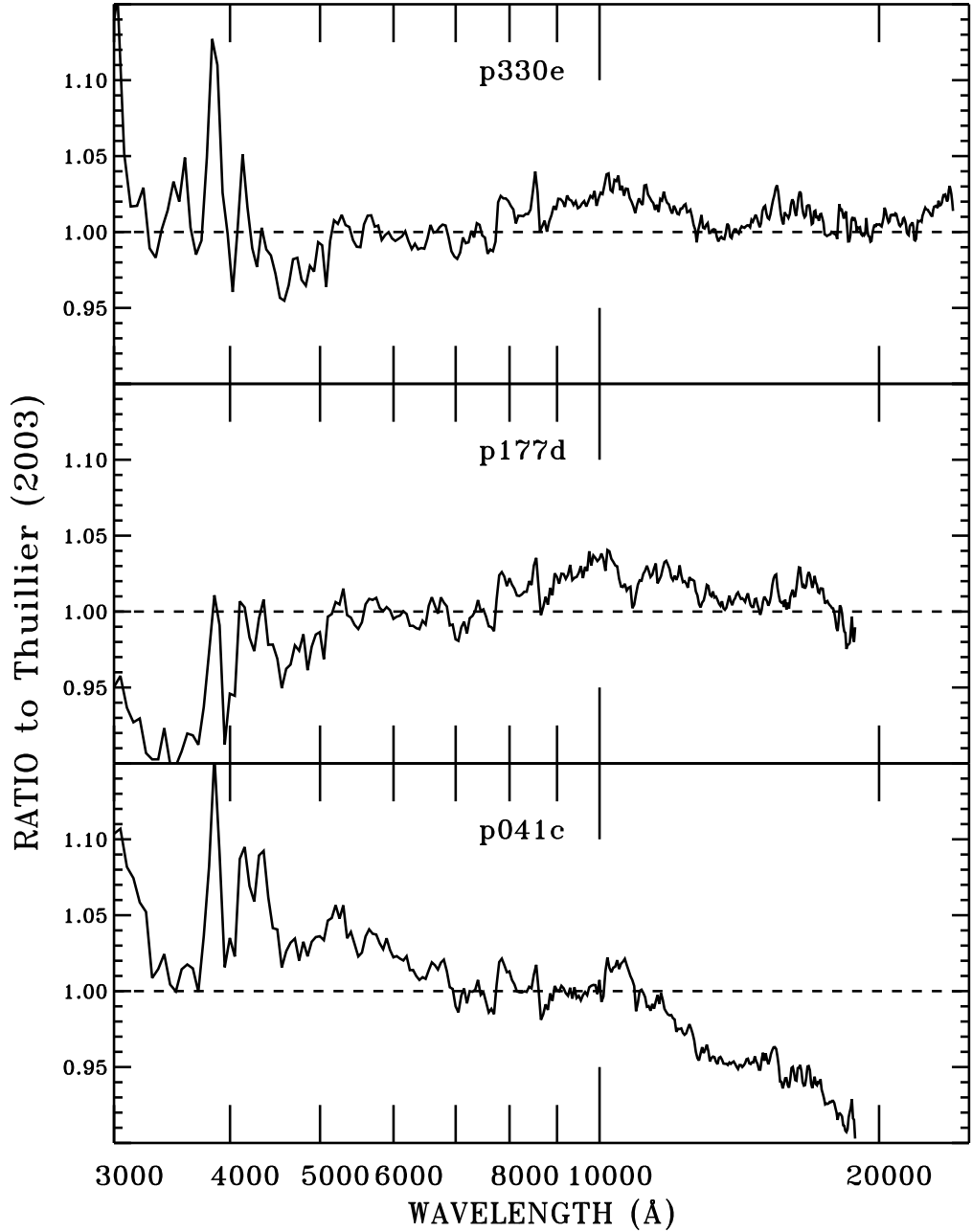}
\caption{Ratio in 50~\AA\ bins of the measured stellar flux for the solar
analogs to the solar flux from Th03 after normalizing to unity at
7000--8000~\AA.}
\end{figure}

In the case of P041C, the slope of the ratio in Figure~7 from 4500--19000~\AA\
has the opposite sign from the slope for the other two stars and suggests that
P041C is significantly hotter than the sun. The models in the Kurucz (1993) grid
can be used to predict the change in slope of the continuum, even though the
lines are not modeled with high fidelity. Figure~9 shows the ratio of the
$T_\mathrm{eff}=5750$ and 6000~K models from the 1993 grid, both of which have
$\log g=4.5$. The smoothed version of this ratio as scaled down by 83/250 is
used to correct the P041C spectrum from 5860~K to the solar
$T_\mathrm{eff}=5777$~K before making its quotient that is fairly flat from
4500--12000~\AA\ in Figure 8.
\begin{figure} 
\centering
\includegraphics[width=\textwidth]{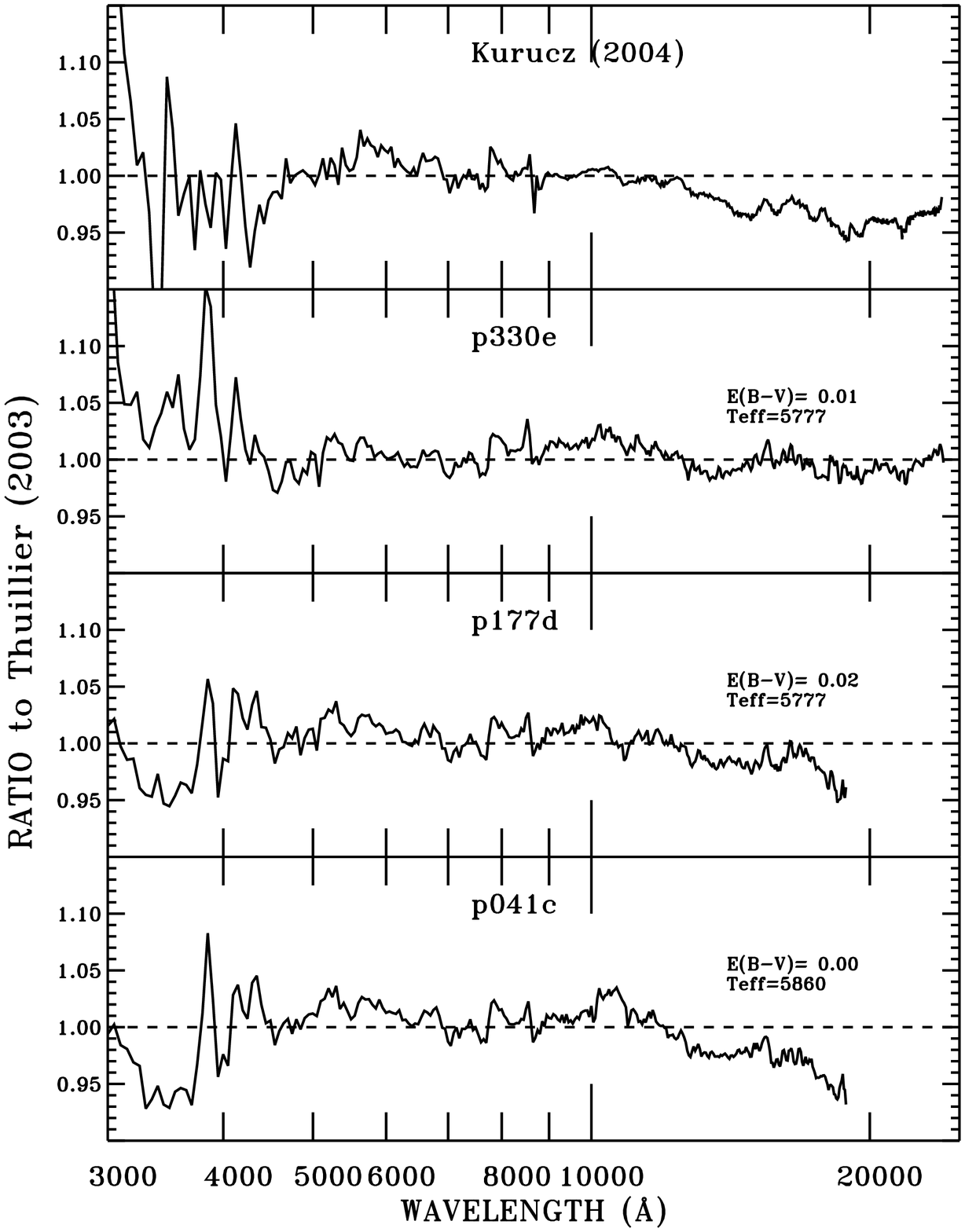}
\caption{Lower three panels: As in Figure~7, except that the solar model has
been adjusted for interstellar reddening, as indicated by E(B$-$V) or for a small
difference in $T_\mathrm{eff}$ from the baseline temperature of 5777~K. Upper
panel: Ratio of $T_\mathrm{eff}=5777$~K Kurucz (2004) solar model to the same
Th03 denominator that is used for all panels in Figures 7--8.}
\end{figure}

Shown in the top panel of Figure~8 is the ratio of the Kurucz (2004) model to
the Th03 measured fluxes. In the IR beyond 1.2~$\mu$m, the Th03 fluxes are
systematically brighter than the high fidelity model by up to 5\%. Price (2004)
points out that the Th03 fluxes are even brighter than earlier 1--2~$\mu$m solar
flux measurements that were already deemed too high by Colina, Bohlin, \&
Castelli (1996). Th03 suggests that this brightening in comparison to the models
is because the sun has  ``...various magnetic structures in the atmosphere....''
Thus, the agreement of P330E with Th03 in the IR suggests that this star has a
similarly high IR flux, while the agreement of P041C with the IR ratio of the
model suggest that P041C is free of such magnetic effect. Apparently, the
effects of magnetic activity in P177D is intermediate between the other two
stars.

\begin{figure} 
\centering
\includegraphics[angle=90,width=.75\textwidth]{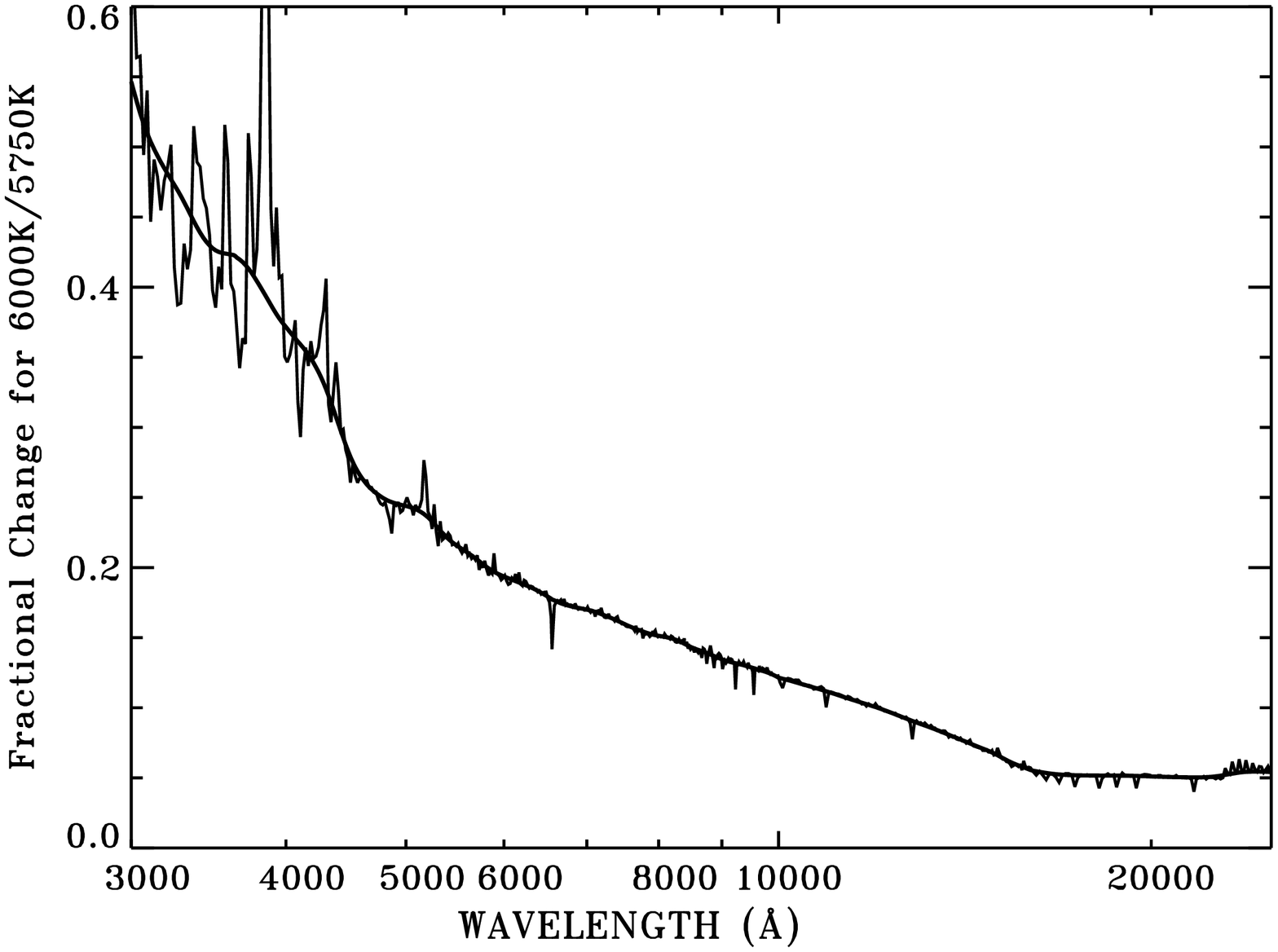}
\caption{Thin line: Fractional change in stellar flux from the ratio of Kurucz
(1993) models for $T_\mathrm{eff}=6000$~K and 5750~K both for $\log g=4.5$.
Thick line: Smooth fit to the thin line to show the expected change in the
continuum for a change of 250~K in the solar model.}
\end{figure}

\begin{figure}  
\centering
\includegraphics[angle=90,width=.95\textwidth]{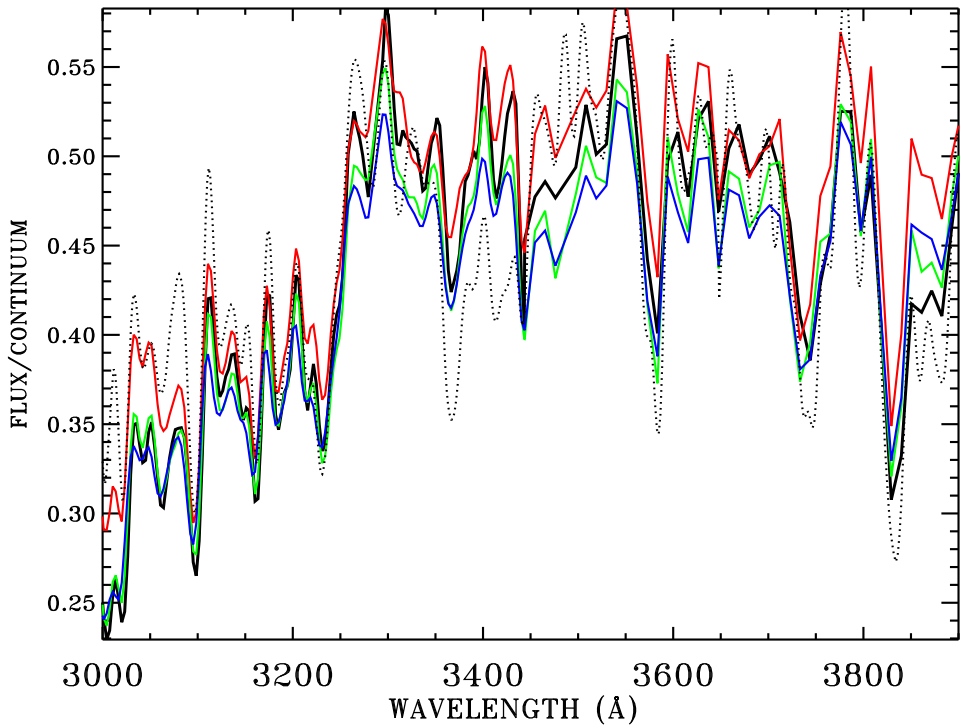}
\caption{Fluxes for the three solar analog stars, the Th03 solar flux, and the
Kurucz (2004) model flux, all of which are divided by the same Kurucz (2004)
continuum level. P330E--red, P177D--Green, P041C--blue, Th03--black, and
Kurucz--dotted.}
\end{figure}

Figure~10 is constructed to understand the deviations from unity at the shorter
wavelengths in Figure~8. To intercompare SEDs from the stars, the sun, and the 
2004 model, all fluxes are corrected for small amounts of interstellar reddening
or temperature differences as in Figure~8, normalized to the model at
7000--8000~\AA, and then divided by the model continuum level, conveniently
supplied along with the computed solar SED (Kurucz 2004). First of all, Figure~10 
demonstrates that there is no systematic broadband calibration error or error
in the average broad band computed stellar flux level, because the stars, the
model, and the solar fluxes agree on average. Second, there is an excellent
agreement in the detailed structure of all five spectra, which demonstrates the
negligible noise level of the data and that the lines of the same chemical
elements are present in the sun, the stars, and in the model. However, large
differences exist in the strength of these spectra features. In particular, the
(Kurucz 2004) model differs from the observations by more than 10\% in some
regions, e.g., at 3360--3440~\AA.
These model differences are due to the poor and incomplete atomic line data and
to elemental abundance uncertainties, which makes ``Matching [theoretical models
to] observed spectra ... hopeless, at least in the near future." (Kurucz 2005b).

Even if solar analog standards cannot be made from computed models, there is the
possibility of using the observed solar SED for all or parts of solar analog
flux distributions. However in a region of heavy line blanketing, Figure~10
illustrates in detail why the 3000--3900~\AA\ region differs so much from unity
for the three stars in Figure~8. While the blue and green traces for P041C and
P177D track each other to better than $\sim$5\% everywhere in Figure~10 and to
better than $\sim$2\% on average, as expected from Figure~8, there are larger
differences from P330E (red) and from the sun (black). Th03 quotes uncertainties
of 2--3\% for the solar flux. Relative errors among the stellar fluxes should be
$<$2\%, because both STIS and FOS have observed P041C, and those independently
determined fluxes agree to $<$2\% in 100~\AA\ bins. The solar trace in Figure~10
ranges from $\sim$5\%, i.e., 0.42/0.44, lower than any star at 3860~\AA\ to about
as high as P330E in the 3300--3360~\AA\ region. 

An extension of the stellar flux distributions is required longward of the
NICMOS limits of 1.9~$\mu$m for P041C and P177D and beyond 2.5~$\mu$m for P330E,
in order to compare with NICMOS photometry. The Th03 solar data extends only to
2.4~$\mu$m, so that the 2004 Kurucz model must be used for the extensions. The
model is normalized to the three stellar observations near the end of their
respective wavelength ranges and adjusted for reddening or $T_\mathrm{eff}$ as
in Figure~8. However, the model fluxes fit the observed IR SEDs poorly, except
for P041C, so that additional errors of $\sim$5\% may be present in the
composite extensions that are constructed out to 3~$\mu$m.\footnote{These are available at
http://www.stsci.edu/hst/observatory/cdbs/calspec.html.}

To summarize, the solar analog method may suffice to form crude standards at the
$\sim$10\% level; but G~type stars with matching visible spectra can differ by a
few percent, especially in regions of heavy line blanketing below
$\sim$4500~\AA. In the IR, intrinsic differences in the magnetic regions
apparently limits the accuracy of the technique to $\sim$5\%. Other types of
stars are more amenable to extrapolation of the observed SEDs outside of the
observed regions, e.g., compare the residuals for faint WDs in Figure~3.

\section{Cohen A- and K Stars}

Four main sequence A-stars and four K~giants have been observed by NICMOS to
provide additional flux standards for \textit{JWST} calibration. Reach et~al.\ (2005) use
these eight stars as calibrators for the Spitzer IRAC, while Cohen et~al.\ 
(2003) describe their provenance. In Figure~11, the NICMOS fluxes are ratioed to
these standard fluxes provided by M.~Cohen (private communication). In the
relevant spectral range below 2.5~$\mu$m, the Cohen K~giant templates are based
on Pickles (1998) library of observed spectra, supplemented by additional
airborne observations longward of 1.22~$\mu$m, as necessary, to fill gaps and
to replace specious blackbody segments. The Cohen dwarf A-star templates are
based on the Kurucz (1993) models with solar abundances.

\begin{figure}  
\centering
\includegraphics[width=\textwidth]{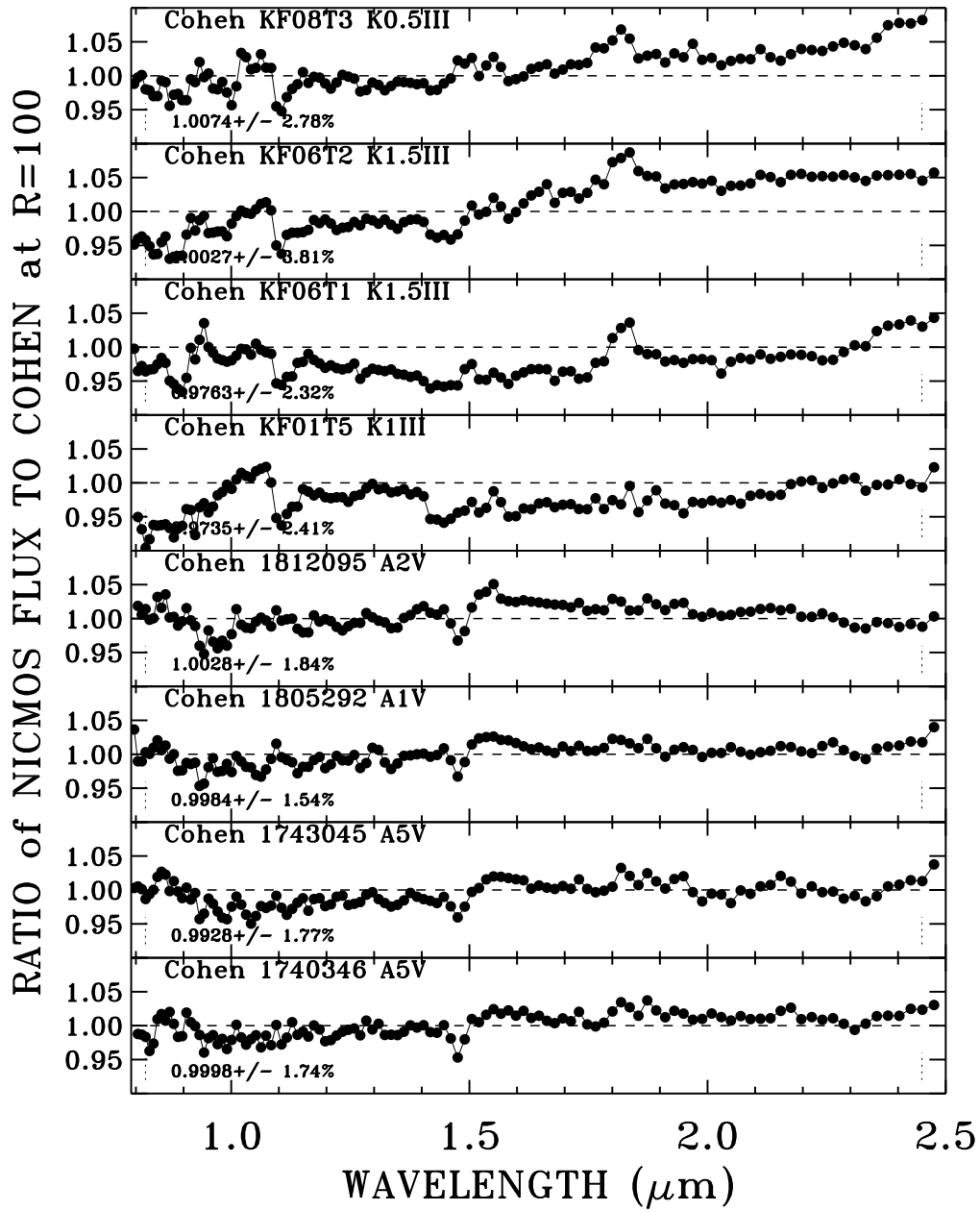}
\caption{Ratio of the NICMOS SEDs to the Cohen template spectra for four main
sequence A-stars and four K~giants. The mean and rms scatter of the ratios
between 0.82 and 2.45~$\mu$m, as delineated by the vertical dotted lines, are
written in each panel. The rms differences at $R=100$ are all below 2\% for the 
A-stars and greater than 2\% for the K~giants, confirming the choice of the
A-stars as the better basis for the calibration of IRAC on the \textit{Spitzer Space
Telescope} (Reach et~al.\ 2005).}
\end{figure}

Figure~11 demonstrates better agreement between the NICMOS and the Cohen 
A-stars, than between NICMOS and the K~giants. Agreement for the A-stars is
typically within 2\% over 0.2~$\mu$m wide bands, with an rms that is always less
than 2\%. Differences between the two sources of K~giant fluxes are often
$\sim$5\% over extensive blocks of wavelength coverage, e.g., the KF06T2 ratio
averages more than 1.05 from 1.8--2.5~$\mu$m, while the rms of all four ratios is
always greater than 2\%. The excellent agreement between these two independent
IR flux calibrations for the A-stars suggests that the WD models that are the
basis for the NICMOS fluxes have a comparable fidelity with the Kurucz (1993)
A-star models that determine the  Cohen A-star SEDs. In contrast, the poorer
agreement for the K~giants may reflect the intrinsic cosmic scatter, e.g., the
strengths of the stellar CO bands in the IR may vary substantially within the
same K-giant sub-type. Reach et~al.\ (2005) also noticed that ``...the K~giants
are systematically offset from the A-stars...'' in the IRAC calibration data and
made a good decision to adopt an A-star only absolute calibration.

\begin{figure}  
\centering
\includegraphics[width=\textwidth]{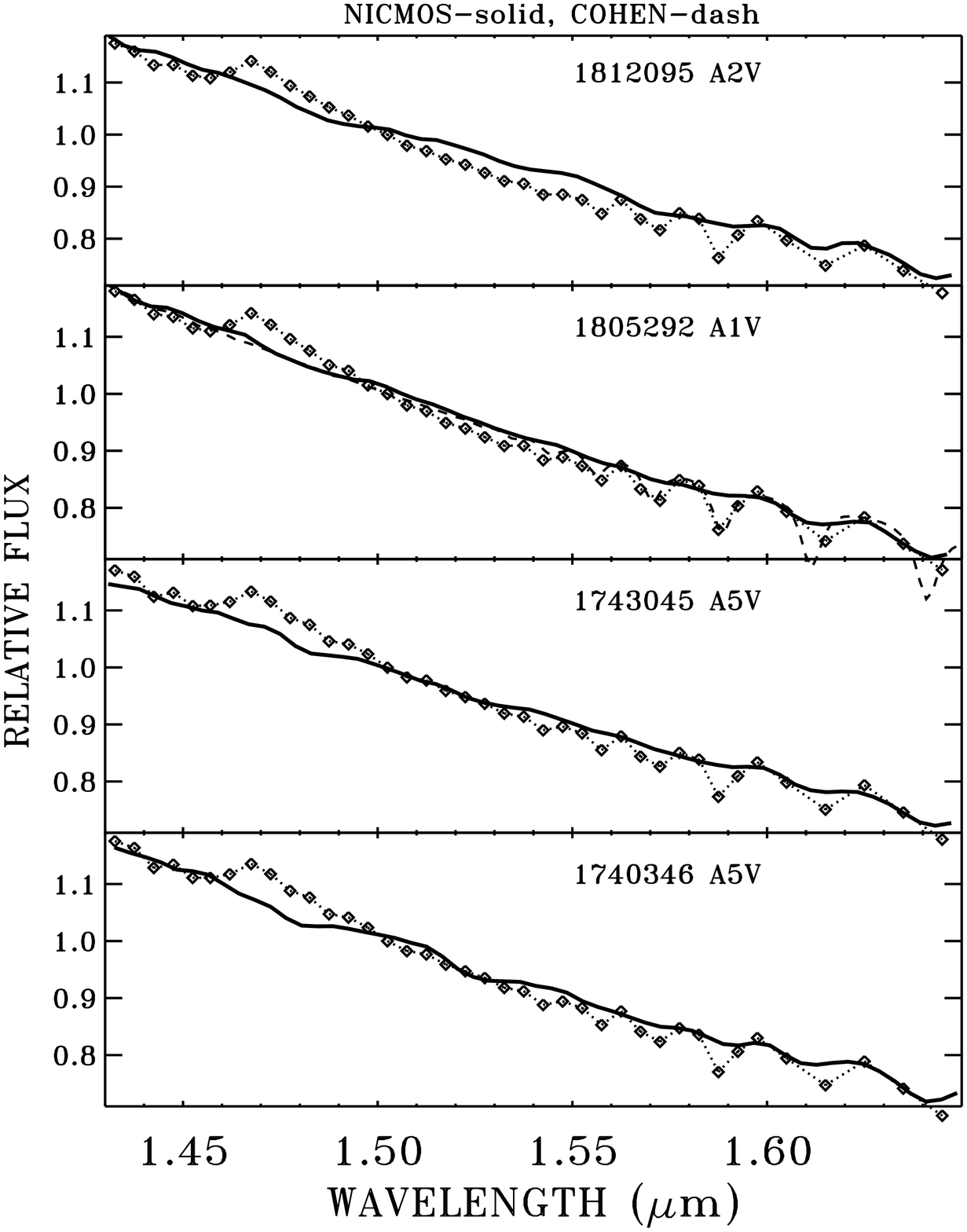}
\caption{Detail of fluxes for A-stars in the region of the biggest systematic
difference between Cohen and NICMOS near 1.5~$\mu$m. Solid lines--NICMOS fluxes,
Dotted lines and diamonds--Cohen template spectra. The differences of a few
percent peak-to-peak are probably due to inaccuracies in Cohen's model flux
templates at the convergence of the Brackett lines to the Brackett continuum at
1.46~$\mu$m. Dashed line in the A1V star panel--Kurucz (2005a) Vega spectrum at
a resolution of $R=500$.}
\end{figure}

Despite the excellent agreement between NICMOS and the Cohen A-stars, there are
some systematic differences that exceed the 1\% accuracy goal. For example, the 
$\sim$4\% dip at 1.47~$\mu$m followed by a 2--4\% bump in the 1.5--1.6~$\mu$m
region is a signature of all four A-star ratios in Figure~11. In Figure~12, all
four NICMOS A-stars are rather smooth across the Brackett jump at 1.46~$\mu$m,
while the A-star templates show two points (diamonds connected with a dotted
line) with positive slope between 1.46 and 1.47~$\mu$m. Also shown as a dashed
line in the A1V star panel of Figure~12 is the high fidelity Vega (A0V) model of
Kurucz (2005a) normalized to the flux for 1805292. This new model with a higher
density of points does agree with NICMOS SED, so that the dashed line is barely
discernable, except at the longer wavelengths in the Brackett lines where the
higher resolution ($R=500$) makes the lines much deeper than in the low resolution
NICMOS spectra. The sampling of the 1993 Kurucz models is much coarser; and
their sampling over the two Brackett lines near 1.61 and 1.64~$\mu$m misses most
of the absorption equivalent width. The old Kurucz (1993) models are not
appropriate for use in regions of spectral features where the goal is to achieve
a $\sim$1\% precision in a SED. Until better models are available, the Cohen
fluxes are adopted for extending the NICMOS fluxes below 0.8~$\mu$m and above
2.5~$\mu$m in the spectra.\footnote{Available at
http://www.stsci.edu/hst/observatory/cdbs/calspec.html.}

\section{Future}

To improve the CALSPEC primary WD standards, a review of the $T_\mathrm{eff}$
and $\log g$ for the NLTE models that best fit the observation is required. An
independent comparison of primary standard star fluxes against laboratory
irradiance standards is an even more fundamental experiment that is probably
best performed by a few sounding rocket flights. See the paper by Mary Beth
Kaiser in this conference proceedings.

In terms of improving the observations of secondary CALSPEC standards relative
to the primary pure hydrogen WD stars, the emphasis of the \textit{HST} program will be
on NICMOS spectrophotometry, because STIS is no longer operational, because the
IR is relatively unexplored by \textit{HST}, and because IR flux standards are required
for \textit{JWST} and potential cosmological dark energy missions, like SNAP. More
observations of the primary WDs and of the secondaries will minimize the rms
NICMOS transfer errors. Now that the techniques for reducing the longest
wavelength G206 grism data with its high background have been proven, more of
the secondary standards should be observed in this 1.9--2.5~$\mu$m range.

\acknowledgments 
Special thanks to R.~Kurucz for computing the high fidelity
$T_\mathrm{eff}=9400$~K model SED for Vega and for comments on a draft of
this paper. M.~Cohen provided his SEDs for the dwarf~A and K~giants. J.~Rhoads
organized the \textit{HST} observing time for the dwarf~A and K~giants, while 
R.~Diaz-Miller submitted the \textit{HST} phase~2 observing proposal. Primary support for
this work was provided by NASA through  the Space Telescope Science Institute,
which is operated by AURA, Inc., under NASA contract NAS5-26555. Additional
support came from DOE through contract number C3691 from the University of
California/Lawrence Berkeley National Laboratory.

\newpage
\question{Freudling} The final absolute accuracy of the calibration of NICMOS
spectra in your presentation was 2 to 3\%. The relative calibration
uncertainties for NICMOS spectra extracted from different parts of the detector
are typically significantly larger than this number due to flat fielding
uncertainties, pixel response function, and non-linearities. To what kind of
observations does the absolute calibration apply? 

\answer{Bohlin} Corrections for the pixel response function due to interpixel
gaps effect the extracted four-pixel-high spectra by as much as $\pm$5\% (Bohlin
et~al.\ 2005); but errors in this correction are $<$1\%. I claim to correct for
non-linearities to $\sim$1\% (Bohlin, Riess, \& de Jong 2006).

My sensitivity calibrations for the NICMOS grism spectra are for a small
region in the lower left hand quadrant of the detector. Figure~3 shows residuals of
$<$1\% for the three primary standards for this region of the detector.
An early post-cryo-cooler observation of P330E in 2002 by Roger Thompson
is an average of 15 different positions over the entire NIC-3 detector.
The whole detector average and my spectrum of P330E are consistent to 2\%;
however, flat fielding errors of 3\% at any single location on NIC3 would not
be surprising. A new grism flat field data cube should be obtained.

\question{Adelman}  1)~Although it is possible to fit the continuum of Vega
over a fair range of wavelengths with the predictions of a single Kurucz model,
it is not possible to fit the line spectrum without modeling the changes of
temperature and gravity from the pole to the equator. Synthetic Vega spectra
depend on factors such as the difference of the equatorial and polar
temperatures, the inclination angle of the rotational axis and the equatorial
velocity.

2)~Besides problems due to atomic and molecular data, synthetic spectra of solar
like stars longward of the Balmer confluence also change with the adopted theory
of convection. What is needed to get better matches is a theory of convection
which is based on real physics and physically defined boundaries. Simplified
models such as mixing-length theory and its variants are bound to be inadequate.

\answer{Bohlin}
 1)~The measured continuum flux of Vega from 3200--10000~\AA\
fits a single temperature Kurucz model at 9400~K to 2\% and to 1\% over most of
this region (see Figure~5). The largest differences occur in the
psuedo-continuum from 3650--4450~\AA\ in the Balmer line confluence, where the
exact physics of the hydrogen atom is still problematic. What Saul says is
certainly true for the line cores at high spectral resolution.

2)~No comment.


\newpage
\begin{thebibliography}{}

\bibitem[]{}
Aufdenberg, et al. 2006, Astro-ph, 0603327
\bibitem[]{}
Barstow, M., Good, S., Burleigh, M., Hubeny, I., Holberg, J., \& Levan, A.
	2003, \hbox{MNRAS}, 344, 562
\bibitem[]{}
Bohlin, R.~C. 2000, AJ, 120, 437
\bibitem[]{}
Bohlin, R. 2003, 2002 HST Calibration Workshop, eds.\ S.~Arribas, A.~Koekemoer,
and B.~Whitmore (Baltimore: STScI), p.~115
\bibitem[]{}
Bohlin, R.~C., Dickinson, M.~E., \& Calzetti, D. 2001, AJ, 122, 2118
\bibitem[]{BG}
Bohlin, R.~C., \& Gilliland, R.~L. 2004, AJ, 127, 3508 (BG)
\bibitem[]{}
Bohlin, R.~C., Lindler, D., \& Riess, A. 2005, Instrument Science Report, 
NICMOS 2005-002, (Baltimore: STScI)\footnote{These STScI internal documents can be found at one of these URLs:\\
http://www.stsci.edu/hst/acs/documents/isrs/, http://www.stsci.edu/hst/stis/documents/isrs/, http://www.stsci.edu/hst/nicmos/documents/isrs/.}
\bibitem[]{}
Bohlin, R.~C., Riess, A., \& de Jong, R. 2006, Instrument Science Report, 
NICMOS 2006-002, (Baltimore: STScI)$^5$
\bibitem[]{}
Cohen, M., Megeath, S.~T., Hammersley, P.~L., Martin-Luis, F., \& Stauffer,
J. 2003, AJ, 125, 2645
\bibitem[]{}
Cohen, M., Walker, R.~G., Barlow, M.~J., \& Deacon, J.~R. 1992, AJ, 104, 1650
\bibitem[]{BG}
Colina L., Bohlin, R.~C., \& Castelli, F. 1996, AJ, 112, 307
\bibitem[]{}
Finley, D.~S., Koester, D., \& Basri, G. 1997, ApJ, 488, 375
\bibitem[]{}
Goudfrooij, P., \& Bohlin, R. 2006, Instrument Science Report, STIS 2006-03,
(Baltimore: STScI)$^5$
\bibitem[]{}
Hayes, D.~S. 1985, in Calibration of Fundamental Stellar
	Quantities, Proc.\ of IAU Symposium No.~111, eds.\ D.~S.\ Hayes, L.~E.\
	Pasinetti, A.~G.\ Davis Philip (Reidel: Dordrecht), p.~225
\bibitem[]{}
Hubeny, I., \& Lanz, T. 1995, ApJ, 439, 875
\bibitem[]{}
Kurucz, R.~L. 1993, CD-ROM 13, ATLAS9 Stellar Atmosphere Programs and 2km/s Grid
(Cambridge: SAO)
\bibitem[]{}
Kurucz, R.~L. 2004, http://kurucz.harvard.edu/
\bibitem[]{}
Kurucz, R.~L. 2005a, Vega spectrum at $T_\mathrm{eff}=9400$~K, $\log g=3.90$,
	[M/H]~$=-0.5$, and zero microturbulent velocity, personal communication
\bibitem[]{}
Kurucz, R.~L. 2005b,  New atlases for solar flux, irradiance, central intensity,
     and limb intensity. Presented at the workshop ``ATLAS12 and Related
      Codes'' held in Trieste 11--15 July, 2005, Memorie della Socienta 
     Astronomica Italiana Supplementi, vol.~8, pp.~158--160
\bibitem[]{}
Landolt, A.~U. 1992, AJ, 104 340
\bibitem[]{}
Megessier, C. 1995, A\&A, 296, 771
\bibitem[]{}
Peterson, D.~M., et al. 2006, Nature, 440, 896
\bibitem[]{}
Pickles, A. 1998, PASP, 110, 863
\bibitem[]{}
Price, S.~D. 2004, Sp.\ Sci.\ Rev., 113, 409
\bibitem[]{}
Reach, W.~T., et al. 2005, PASP, 117, 978
\bibitem[]{}
Rottman, G., Harder, J., Fontenla, J., Woods, T., White, O., \& Lawrence, G.
2005, Sol.\ Phys., 230, 205
\bibitem[]{}
Savage, B., \& Mathis, J. 1979, ARAA, 17, 73
\bibitem[]{}
Stys, D.~J., Bohlin, R.~C., \& Goudfrooij, P. 2004, Instrument Science Report,
STIS 2004-04, (Baltimore:ST ScI)$^5$
\bibitem[]{Th03}
Thuillier, G., Herse, M., Labs, D., Foujols, T., Peetermans, W., Gillotay, D.,
Simon, P., \& Mandel, H. 2003, Sol.\ Phys., 214, 1 (Th03)
\end{thebibliography}
\end{document}